\documentclass[letter,twocolumn]{jpsj2} 
\title{Superconducting Volume Fraction in Overdoped Regime of La$_{2-x}$Sr$_x$CuO$_4$: Implication for Phase Separation from Magnetic-Susceptibility Measurement}

\author{\textsc{Yoichi Tanabe}, \textsc{Tadashi Adachi}\thanks{E-mail address: adachi@teion.apph.tohoku.ac.jp}, \textsc{Takashi Noji} and \textsc{Yoji Koike}}

\inst{Department of Applied Physics, Graduate School of Engineering, Tohoku University, \\6-6-05, Aoba, Aramaki, Aoba-ku, Sendai 980-8579}

\abst{We have grown a single crystal of La$_{2-x}$Sr$_x$CuO$_4$ in which the Sr concentration, $x$, continuously changes from 0.24 to 0.29 in the overdoped regime and obtained many pieces of single crystals with different $x$ values by slicing the single crystal. 
From detailed measurements of the magnetic susceptibility, $\chi$, of each piece, it has been found that the absolute value of $\chi$ at the measured lowest temperature 2 K, $|\chi_{\rm 2 K}|$, on field cooling rapidly decreases with increasing $x$ as well as the superconducting (SC) transition temperature. 
As the value of $|\chi_{\rm 2 K}|$ is regarded as corresponding to the SC volume fraction in a sample, it has been concluded that a phase separation into SC and normal-state regions occurs in a sample of La$_{2-x}$Sr$_x$CuO$_4$ in the overdoped regime.}

\kword{phase separation, La$_{2-x}$Sr$_x$CuO$_4$, magnetic susceptibility, Sr-concentration-gradient single crystal, superconducting volume fraction}

\begin{document}
\maketitle

The study of the electronic states of high-$T_{\rm c}$ cuprates has been a central issue to elucidate the mechanism of high-$T_{\rm c}$ superconductivity. 
Compared with the underdoped regime in which extensive studies have been carried out, physical properties in the overdoped regime have not been so clarified. 
The microscopic electronic inhomogeneity in the CuO$_2$ plane suggested from scanning-tunneling-microscopy (STM) measurements was a sensational observation.~\cite{pan} 
In optimally doped Bi$_2$Sr$_2$CaCu$_2$O$_{8+\delta}$ (Bi-2212), it was suggested that both the local density of states and the superconducting (SC) gap were spatially inhomogeneous. 
Such electronic inhomogeneity was observed in slightly underdoped and overdoped Bi-2212 as well.~\cite{lang} 
However, the inhomogeneity in the CuO$_2$ plane has still been a controversial issue, because STM measurements are very sensitive to the surface electronic state, which may sometimes be rather different from the bulk one. 
In order to confirm the inhomogeneity, measurements reflecting bulk properties of a sample are absolutely required. 

As for the electronic inhomogeneity in the overdoped regime, transverse-field muon-spin-relaxation ($\mu$SR) measurements in the overdoped regime of Tl$_2$Ba$_2$CuO$_{6+\delta}$ revealed that the muon-spin depolarization rate in the ground state, regarded as being proportional to the SC carrier density divided by the effective mass, $n_{\rm s}/m^{\rm *}$, decreased with an increase in the hole concentration, $p$.~\cite{uemura,niedermayer} 
It was speculated that a microscopic phase separation into SC and normal-state regions occurred in the overdoped high-$T_{\rm c}$ cuprates.~\cite{uemura2} 
Similar results were also obtained from the $\mu$SR measurements in the overdoped regime of (Y, Ca)Ba$_2$Cu$_3$O$_{7-\delta}$ (Y/Ca-123) and Tl$_{1-y}$Pb$_y$Sr$_2$Ca$_{1-y}$Y$_y$Cu$_2$O$_7$.~\cite{bernhard}

One conclusive way to prove the phase separation is to study the SC volume fraction. 
Formerly, Nagano {\it et al}. reported from measurements of the magnetic susceptibility, $\chi$, in La$_{2-x}$Sr$_x$CuO$_4$ (LSCO) that the shielding volume fraction, obtained from $\chi$ at 4.2 K on zero-field cooling using a bulk sample, was almost 100\% in the range $0.07 \le x \le 0.27$.~\cite{nagano} 
They insisted that the SC volume fraction was nearly 100\% in the whole SC regime of LSCO. 
In general, however, the shielding volume fraction is larger than the real SC volume fraction in the case of the existence of non-SC regions in a sample, namely, when the microscopic phase separation into SC and normal-state regions takes place. 
Therefore, the so-called Meissner volume fraction obtained from $\chi$ on field cooling is necessary for the estimation of the SC volume fraction. 

In this paper, we have investigated the detailed $p$ dependence of $\chi$ on field cooling, using many pieces of single crystals with different $x$ values obtained by slicing a large Sr-concentration-gradient single crystal in which the Sr concentration continuously changes from 0.24 to 0.29 in the overdoped regime of LSCO. 
Our study is based upon precise measurements using these adequate samples, because the samples obtained from the Sr-concentration-gradient single crystal have the same crystallinity, which makes the vortex-pinning effect by crystal imperfections almost identical to each other. 
The other reason is that the powder samples prepared by crushing each piece of single crystal are used in order to keep the influence of the size of the sample, the demagnetizing field and the anisotropy of $\chi$ almost identical to each other. 
We have found that not only the SC transition temperature, $T_{\rm c}$, but also the absolute value of $\chi$ at the measured lowest temperature 2 K, $|\chi_{\rm 2 K}|$, rapidly decreases with increasing $x$, indicating the decrease in the SC volume fraction with increasing $x$. 
These findings strongly suggest that a phase separation into SC and normal-state regions occurs in the overdoped regime of LSCO. 

A Sr-concentration-gradient single crystal of LSCO with $x$ ranging from 0.24 to 0.29 was grown by the traveling-solvent floating-zone (TSFZ) method under flowing O$_{\rm 2}$ gas of 4 or 9 bar. 
The details of the preparation of powders for the feed and solvent rods have been reported elsewhere.~\cite{kawamata} 
For the feed rod, stoichiometric powders of LSCO with $x=0.25$, 0.26, 0.27, 0.28, 0.29, 0.30 were prepared. 
Then, the obtained fine powders of each $x$ were placed into a thin-walled rubber tube in order and formed into a cylindrical rod under hydrostatic pressure. 
That is, $x$ was changed discretely from 0.25 to 0.30 at intervals of 0.01 in the feed rod in which the length of each $x$ was $\sim$ 20 mm. 
For the solvent, the composition was in the molar ratio of La : Sr : Cu = 3 : 3 : 7. 
The as-grown single-crystal rod was annealed in flowing O$_{\rm 2}$ gas of 1 bar at 900$^{\rm o}$C for 50 h, cooled down to 500$^{\rm o}$C at a rate of 8$^{\rm o}$C/h, kept at 500$^{\rm o}$C for 50 h and then cooled down to room temperature at a rate of 8$^{\rm o}$C/h. (We call this 900$^{\rm o}$C annealing.) 
In order to fill up remaining oxygen vacancies after the 900$^{\rm o}$C annealing, the crystal was further annealed in high-pressure O$_{\rm 2}$ gas of 1000 bar at 500$^{\rm o}$C for 2 weeks. (We call this HP annealing.) 
The Sr content of each part of the single-crystal rod was analyzed by inductively coupled plasma atomic emission spectrometry (ICP-AES). 
The oxygen deficiency was estimated from iodometric titration. 
$\chi$ measurements were carried out both on zero-field cooling and on field cooling at low temperatures down to 2 K, using a superconducting quantum interference device (SQUID) magnetometer. 
The powder samples, obtained by slicing the single-crystal rod perpendicularly to the direction of the crystal growth to form many pieces of 1 mm thickness and crushing them up, were used for the $\chi$ measurements. 
The particle size in the powder samples was smaller than $\sim40$ $\mu$m.
The distribution of $x$ in one powder sample, $\Delta x$, was estimated from the ICP-AES to be $\sim 0.0005$.

\begin{figure}[tbp]
\begin{center}
\includegraphics[width=0.85\linewidth]{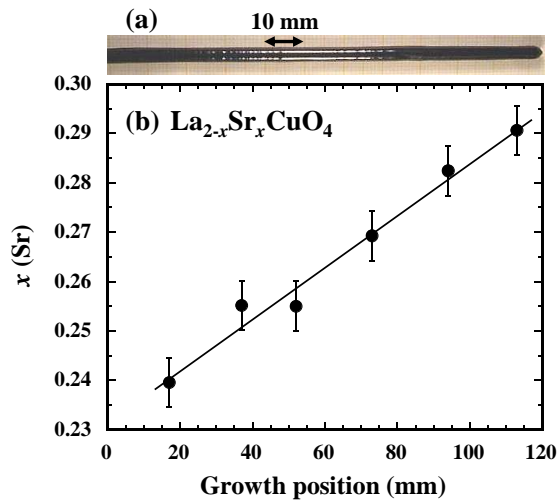}
\end{center}
\caption{(a) (Color online) As-grown single-crystal rod of Sr-concentration-gradient La$_{2-x}$Sr$_x$CuO$_4$. (b) Sr-concentration, $x$, estimated from ICP-AES vs growth position of single-crystal rod. Note that the horizontal axis corresponds to the photo in (a).}  
\label{icp} 
\end{figure}

Figure \ref{icp}(a) displays an as-grown single-crystal rod of the Sr-concentration-gradient LSCO. 
The size is 4 mm in diameter and 120 mm in length. 
The quality was checked by X-ray back-Laue photography to be good. 
The full-width at half maximum of the (006) rocking curve in a part of the crystal was measured to be $\sim 0.1^{\rm o}$, which is comparable to that in the previous report.~\cite{komiya}
The crystal was also checked by powder X-ray diffraction. 
Bragg peaks of LSCO and no impurities could be observed. 
As shown in Fig. \ref{icp}(b), the $x$ estimated from the ICP-AES increases almost monotonically in the direction of the crystal growth. 
Therefore, it is concluded that the growth of the Sr-concentration-gradient single crystal has been successful as we expected. 

The oxygen deficiency, $\delta$, in La$_{2-x}$Sr$_x$CuO$_{4-\delta}$ with $x=0.271$ grown under flowing O$_2$ gas of 4 bar was estimated to be $\delta = 0.014 \pm 0.01$ after the 900$^{\rm o}$C annealing and changed to be $\delta = 0.008 \pm 0.01$ after the following HP annealing. 
Therefore, the HP annealing is regarded to cause an almost complete filling up of oxygen vacancies in the overdoped regime of La$_{2-x}$Sr$_x$CuO$_{4-\delta}$ of $x$ at least below $\sim$ 0.27.~\cite{oxygen} 

\begin{figure*}[tb]
\begin{center}
\includegraphics[width=0.8\linewidth]{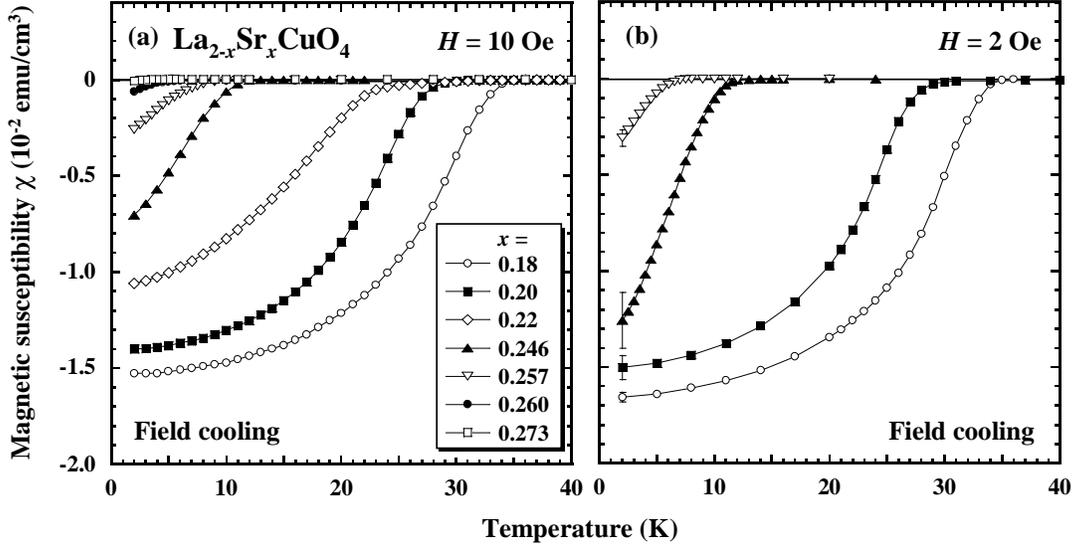}
\end{center}
\caption{Temperature dependence of magnetic susceptibility, $\chi$, for La$_{2-x}$Sr$_x$CuO$_4$ with $x = 0.246 - 0.273$ in magnetic field of (a) 10 Oe and (b) 2 Oe on field cooling. The data for the polycrystalline powder samples with $x=0.18 - 0.22$ are also plotted for reference.~\cite{oki} Note in (b) that error bars of each $x$ shown at 2 K for reference were estimated from the dispersion of the magnetization curve in low fields.}
\label{chi} 
\end{figure*}

The temperature dependence of $\chi$ in a magnetic field of 10 Oe on field cooling for LSCO with $x = 0.246 - 0.273$ is shown in Fig. \ref{chi}(a), together with the data for polycrystalline powder samples of LSCO with $x = 0.18 - 0.22$.~\cite{oki} 
The SC transition width appears to be rather broad for the samples shown here, which is probably due to the temperature dependence of the penetration depth. 
With increasing $x$, $T_{\rm c}$ decreases and disappears for $x \ge 0.273$. 
Moreover, $|\chi_{\rm 2 K}|$ also decreases with increasing $x$ and becomes almost zero for $x \ge 0.273$. 
Compared with the behavior of $\chi$ for $x = 0.18 - 0.22$, $\chi$ for $x \ge 0.246$ is not saturated even at 2 K but seems to continue decreasing with temperature. 
However, the estimation of the extrapolated values of $\chi$ at 0 K allows us to find that $|\chi|$ in the ground state, namely, the Meissner volume fraction decreases with increasing $x$ for $x \ge 0.246$. 

Here, we mention the relation between $|\chi|$ on field cooling and the magnitude of the applied magnetic field, $H$. 
When $H$ is over the lower critical field, $H_{\rm c1}$, or when the vortex-pinning effect is strong, $|\chi|$ tends to decrease with increasing $H$. 
Therefore, we also measured $\chi$ for several $x$ values on field cooling in a low magnetic field of 2 Oe. 
As shown in Fig. \ref{chi}(b), $|\chi_{\rm 2 K}|$ apparently decreases with increasing $x$, indicating that the decrease in the Meissner volume fraction with increasing $x$ is not due to $H$. 
Accordingly, these results suggest that the Meissner volume fraction decreases inherently with increasing $x$ in the overdoped regime of LSCO. 

\begin{figure}[bp]
\begin{center}
\includegraphics[width=0.9\linewidth]{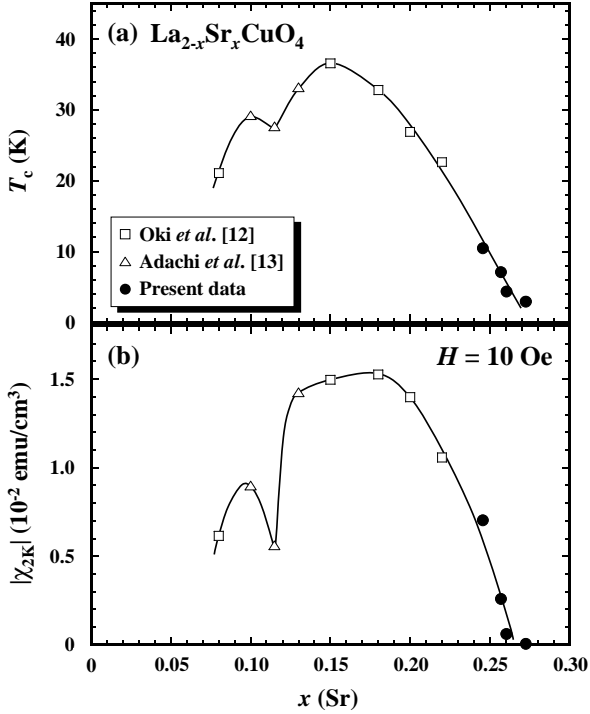}
\caption{Sr-concentration, $x$, dependence of (a) $T_{\rm c}$ defined as cross point between extrapolated line of steepest part of Meissner diamagnetism and zero susceptibility and (b) absolute value of $\chi$ at 2 K, $|\chi_{\rm 2 K}|$, of La$_{2-x}$Sr$_x$CuO$_4$. The data for the polycrystalline powder samples with $x = 0.08 - 0.22$ are also plotted for reference.~\cite{adachi,oki}}
\label{tc-chi}
\end{center}
\end{figure}

Figure \ref{tc-chi} displays the $x$ dependence of $T_{\rm c}$, defined as the cross point between the extrapolated line of the steepest part of the Meissner diamagnetism and zero susceptibility, and $|\chi_{\rm 2 K}|$ in $H=10$ Oe of LSCO. 
The data for the polycrystalline powder samples with $x=0.08 - 0.22$ in $H=10$ Oe on field cooling are also plotted for reference.~\cite{adachi,oki} 
$T_{\rm c}$ decreases with increasing $x$ for $x > 0.15$ and disappears at $x \sim 0.273$, which is consistent with the previous report.~\cite{nagano} 
On the other hand, the $x$ dependence of $|\chi_{\rm 2 K}|$ is quite suggestive; it shows the maximum at $x\sim0.18$, decreases abruptly with increasing $x$ for $x > 0.18$ and becomes zero at $x \sim 0.273$. 
That is, both $T_{\rm c}$ and $|\chi_{\rm 2 K}|$ seem to decrease cooperatively with increasing $x$ in the overdoped regime of LSCO. 

First, we discuss the reason why $|\chi_{\rm 2 K}|$ decreases with increasing $x$ in the overdoped regime of LSCO. 
There are two possibilities. 
One is an increase in the vortex-pinning effect with increasing $x$. 
In general, the vortex pinning is affected by crystal imperfections such as oxygen vacancies.~\cite{okuya} 
The number of oxygen vacancies in a sample tends to increase with $x$ in the overdoped regime of LSCO, so that the vortex-pinning effect increases with $x$. 
In the present study, however, $\delta$ was estimated to be less than 0.01 and almost the same for all samples of La$_{2-x}$Sr$_x$CuO$_{4-\delta}$ with $x = 0.246 - 0.273$. 
Moreover, through the HP annealing for $x=0.273$, $|\chi_{\rm 2 K}|$ negligibly changed. 
Accordingly, it is unlikely that the decrease in $|\chi_{\rm 2 K}|$ with increasing $x$ is due to the increase in the vortex-pinning effect. 

The other is a decrease in the SC volume fraction. 
In fact, a similar decrease in the SC volume fraction was observed from $|\chi_{\rm 2 K}|$ on field cooling in the partially Zn- or Ni-substituted La$_{2-x}$Sr$_x$Cu$_{1-y}$(Zn, Ni)$_y$O$_4$ around $x=0.115$.~\cite{adachi,adachi2} 
It is possible that a phase separation into SC and normal-state regions in which the ground state is regarded as a so-called Fermi-liquid state takes place, so that the SC volume fraction decreases with increasing $x$. 
This is in sharp contrast with the model proposed by Niedermayer {\it et al.}~\cite{niedermayer} that the superfluid and normal quasi-particles are homogeneously distributed in the CuO$_2$ plane in the overdoped regime. 

Next, we discuss the reason why $T_{\rm c}$ decreases with increasing $x$ under the phase-separated state in the overdoped regime of LSCO. 
One possible reason is a decrease in $n_{\rm s}$ in the SC region. 
The screening of the Coulomb interaction between holes in the normal state with $p > 0.19$ per Cu is believed to be weaker than that in the Fermi-liquid state with $p \sim 0.30$ per Cu. 
Therefore, in the case that the loss of the Coulomb energy in the normal-state region surpasses the SC condensation energy in the SC region, it is possible that holes tend to gather into the normal Fermi-liquid state region out of the SC region, leading to the decrease in $n_{\rm s}$ in the SC region.~\cite{uemura2} 
The decrease in $n_{\rm s}$ is expected to result in the decrease in $T_{\rm c}$, as seen in the underdoped and optimally doped high-$T_{\rm c}$ cuprates.~\cite{uemura3} 
Another possible reason is a proximity effect. 
Supposed that, as suggested from the STM measurements,~\cite{pan,lang} a microscopic phase separation into SC and normal-state regions takes place, normal quasi-particles in normal-state regions penetrate into SC regions across the boundary between SC and normal-state regions due to the proximity effect. 
When the size of each SC region decreases with increasing $x$, the development of the SC order parameter in each SC region is suppressed, resulting in the decrease in $T_{\rm c}$. 
Therefore, assuming the microscopic phase separation, both decreases in $T_{\rm c}$ and the SC volume fraction with increasing $x$ can be explained simultaneously. 
Linked with previous STM~\cite{lang} and $\mu$SR~\cite{uemura,niedermayer,uemura2} results, a microscopic phase separation into SC and normal-state regions might be an intrinsic bulk property of the overdoped high-$T_{\rm c}$ cuprates. 

The origin of the phase separation in the overdoped regime of LSCO is an open issue. 
One possible origin is associated with the decrease in the SC condensation energy in the overdoped regime. 
The SC condensation energy tends to decrease with increasing $p$ toward the Fermi-liquid state with $p \sim 0.30$ in the overdoped regime.~\cite{matsuzaki} 
In fact, the change of the Fermi-surface topology from hole-like to electron-like around $x=0.22$ with increasing $x$, observed from an angle-resolved photoemission experiment,~\cite{ino} may result in the decrease in the SC condensation energy in the overdoped regime because of the reduction of the density of states originating from the so-called flat band around ($\pi$, 0) and (0, $\pi$) in the reciprocal lattice space in which the magnitude of the $d$-wave SC gap shows the maximum.
In this case, a phase separation into the SC region with $p \sim 0.19$ in which the SC condensation energy is maximum and the normal Fermi-liquid state region with $p \sim 0.30$ possibly takes place in the CuO$_2$ plane so as to obtain a large gain of the SC condensation energy in the SC region at a small cost of the Coulomb energy in the normal-state region as mentioned in the preceding paragraph. 
Here, it is noted that the inhomogeneity of the electrostatic potential in the CuO$_2$ plane due to the randomness of La/Sr may assist in generating the phase separation.

Another possible origin of the phase separation is the doping of holes not into the O-$2p$ state but directly into the Cu-$3d$ state. 
In the overdoped regime, it is possible that the O-$2p$ band and the so-called lower Hubbard band of the Cu-$3d$ state overlap, so that a part of holes are doped into the Cu-$3d$ band. 
Holes doped into the Cu-$3d_{3z^2-r^2}$ orbital tend to be localized, resulting in the appearance of free Cu spins and pair-breaking around themselves in the CuO$_2$ plane. 
On the other hand, holes doped into the Cu-$3d_{x^2-y^2}$ orbital disturb the antiferromagnetic correlation between Cu spins, possibly relating to the appearance of high-$T_{\rm c}$ superconductivity, more markedly than holes doped into the O-$2p$ orbital and may produce free Cu spins as in the case of the partially Zn-substituted LSCO.~\cite{mahajan} 
In any case, both of them may bring about the local destruction of superconductivity and the formation of a normal-state region around the Cu site with a hole, leading to a microscopic phase separation. 
In fact, this scenario is supported by the following two experimental results. 
The first is the appearance of the Curie term in the temperature dependence of $\chi$ in the overdoped regime of LSCO,~\cite{takagi,oda,oda2} corresponding to the appearance of free Cu spins in the CuO$_2$ plane. 
The second is the suppression of the incommensurate spin correlation in the overdoped regime of LSCO, suggested from the broadening of the incommensurate magnetic peaks in the inelastic neutron-scattering experiment.~\cite{wakimoto} 

Finally, we comment on the $s$-wave SC order parameter mixing into the $d$-wave one in the overdoped regime of Y/Ca-123 and Bi-2212 systems pointed out from Raman-scattering measurements.~\cite{masui} 
Considering the phase-separated state in which the magnitude of the SC gap is spatially inhomogeneous, it is possible that the superconductivity with the $s$-wave SC order parameter mixing into the $d$-wave one appears at the boundary between $d$-wave SC and normal-state regions.~\cite{andreev} 

In summary, from the $\chi$ measurements, not only $T_{\rm c}$ but also the SC volume fraction has been found to decrease rapidly with increasing $x$ in the overdoped regime of LSCO. 
It is concluded that SC and normal-state regions are phase-separated in a sample. 
Linked with the STM~\cite{lang} and $\mu$SR~\cite{uemura,niedermayer,uemura2} results, the microscopic phase separation might be a generic feature of the electronic state in the overdoped high-$T_{\rm c}$ cuprates. 
At present, however, it is unclear whether the phase separation is microscopic or macroscopic, because there are several possible origins to bring about the phase separation in the overdoped regime. 

Fruitful discussions with Prof. M. Kato, Drs. H. Eisaki, T. Nishizaki and H. Tsuchiura are gratefully acknowledged. 
This work was supported by the Iketani Science and Technology Foundation and also by a Grant-in-Aid for Scientific Research from the Ministry of Education, Culture, Sports, Science and Technology, Japan.


\begin{thebibliography}{99}
\bibitem{oxygen}
From the $\chi$ measurement in $x=0.273$, $T_{\rm c}$ was found to increase from below 2 K to 2.9 K through the HP annealing. The enhancement of $T_{\rm c}$ through the decrease in $\delta$ has also been reported in LSCO with $x=0.21$.~\cite{okuya} These suggest that the oxygen deficiency mainly results in the decrease in $T_{\rm c}$ due to the degradation of the SC properties rather than the increase in $T_{\rm c}$ due to the decrease in $p$ in the overdoped regime.
\end{thebibliography}
\end{document}